\documentclass[a4paper,11pt]{article}
\pdfoutput=1 

\usepackage{jinstpub} 

\title{\boldmath Fast DAQ system with image rejection for axion dark matter searches}

\author[a,b]{S. Ahn,}
\author[b]{M. J. Lee,}
\author[a,b]{A. K. Yi,}
\author[a,b,1]{B. Yeo,\note{Now at Department of Physics, University of California, CA 94720, Berkeley, USA}}
\author[b,2]{B. R. Ko,\note{Corresponding author}}
\author[a,b]{and Y. K. Semertzidis}

\affiliation[a]{Department of Physics, Korea Advanced Institute of
  Science and Technology (KAIST), Daejeon 34141, Republic of Korea}
\affiliation[b]{Center for Axion and Precision Physics Research (CAPP),
  Institute for Basic Science (IBS), Daejeon 34051, Republic of Korea}

\emailAdd{brko@ibs.re.kr}

\abstract{
  A fast data acquisition (DAQ) system for axion dark matter searches
  utilizing a microwave resonant cavity, also known as axion haloscope
  searches, has been developed with a two-channel digitizer that can
  sample 16-bit amplitudes at rates up to 180 MSamples/s. First, we
  realized a practical DAQ efficiency of greater than 99\% for a
  single DAQ channel, where the DAQ process includes the online fast
  Fourier transforms (FFTs). Using an IQ mixer and two parallel DAQ
  channels, we then also implemented a software-based image rejection
  without losing the DAQ efficiency. This work extends our continuing
  effort to improve the figure of merit in axion haloscope searches,
  the scanning rate.
}
 
\begin{document}
\maketitle
\flushbottom

\section{Introduction}\label{sec:intro}
A very natural solution to the strong $CP$ problem in the Standard
Model of particle physics (SM)~\cite{strongCP} was proposed by Peccei
and Quinn, who introduced a new global symmetry, PQ
symmetry~\cite{PQ}. The breakdown of the PQ symmetry then results in
the axion~\cite{AXION}. Provided the axion mass is light enough, above
$\mathcal{O}(\mu$eV/$c^2$) according to the original
works~\cite{CDM_LOW} or above $\mathcal{O}$(peV/$c^2$) by more recent
works~\cite{OTHER_AXION_PROD} and below
$\mathcal{O}$(meV/$c^2$)~\cite{SN1987}, the axion becomes one of the
most promising candidates for cold dark matter (CDM).
If the axion turns out to be 100\% of CDM, the SM would be promoted as
a model that would govern 31.7\% of the total Universe energy budget,
according to the precision cosmological measurements and the standard
model of Big Bang cosmology~\cite{PLANCK}.

The axion haloscope search by Sikivie~\cite{sikivie} is the
most sensitive axion dark matter search method to date. By employing
a microwave resonant cavity, the axion signal power will resonate
if the axion mass $m_a$ matches the resonant frequency of the cavity
mode $\nu$, $m_a=h\nu/c^2$. Without knowing the axion mass, it is
therefore obligatory for axion haloscope searches to be able to scan
the resonant frequencies corresponding to possible axion
masses. Accordingly, the scanning rate~\cite{scanrate} for a target
signal to noise ratio ${\rm SNR_{target}}$
\begin{equation}
  \frac{d\nu}{dt}=
  \epsilon_{\rm DAQ}
  \frac{\nu b_a}{Q_L}
  \Big(\frac{\epsilon_{\rm SNR}}{\rm SNR_{target}}\Big)^2
  \Big(\frac{P^{a\gamma\gamma}_a}{P_n}\Big)^2
  \propto\epsilon_{\rm DAQ}\epsilon^2_{\rm SNR}\frac{g^4_{a\gamma\gamma}B^4V^2C^2Q_L}{{\rm SNR^2_{target}} T^2_n}
  \label{EQ:RSCAN}
\end{equation}
is usually referred to as the figure of merit in axion haloscope
searches. In Eq.~(\ref{EQ:RSCAN}), $P^{a\gamma\gamma}_a$ is the axion
signal power proportional to $g^2_{a\gamma\gamma}B^2 VC
Q_L$~\cite{sikivie, scanrate}, where $g_{a\gamma\gamma}$ is the
axion-photon coupling strength, $B$ is the static magnetic field
provided by magnets in the axion haloscopes, $V$ is the cavity volume,
$C$ is the form factor representing the overlap between the electric
field of the cavity mode and the static magnetic field whose general
definition can be found in Ref.~\cite{EMFF_BRKO}, and $Q_L$ is the
loaded quality factor of the cavity mode. $P_n$ is the noise power
proportional to the noise temperature $T_n$ and the axion signal
window $b_a$. $\epsilon_{\rm SNR}$ is the reconstruction efficiency of
the SNR in the axion haloscope search analysis
procedures~\cite{Anal-JHEP} and $\epsilon_{\rm DAQ}$ the DAQ
efficiency.
Although there was no explicit $\epsilon_{\rm DAQ}$ in
Ref.~\cite{scanrate} or in our previous publication~\cite{Anal-JHEP},
the DAQ efficiency is in practice very
significant to the scanning rate, e.g., it was 47\% with the spectrum
analyzer R\&S$^{\tiny\circledR}$FSV4~\cite{RSH} in the CAPP-8TB
experiment~\cite{CAPP-8TB-NIM}.
By doubling the $\epsilon_{\rm DAQ}$, the speed of our axion haloscope
searches will be boosted by a factor of two.
In this work, we have realized a practical DAQ efficiency of greater
than 99\% using a fast digitizer, in a continuing effort following our
previous study~\cite{Anal-JHEP} to improve the scanning rate, where
the DAQ process also includes online FFTs and writing the outputs to
disk.

Axion haloscope searches generally employ a heterodyne receiver which
introduces unwanted image backgrounds, thus could increase the
background level about twice. With an image rejection of 30 dB, the
background level would increase 0.1\% which little affects the
${\rm SNR_{target}}$, thus the scanning rate. The image rejection can
be accomplished with a sharp bandpass filter~\cite{CAPP-8TB-NIM}, an
image rejection mixer~\cite{ADMX-NIM}, or software utilizing in-phase
(I) and quadrature (Q) signals from an IQ mixer~\cite{HAYSTAC-NIM}. We
have implemented a software-based image rejection using an IQ mixer,
and both DAQ channels on the digitizer. Without loss in
$\epsilon_{\rm DAQ}$, we have realized a fast DAQ system equipped with
an image rejection of about 35 dB over a frequency range from 600 to
2200 MHz. Thus, it can be adopted by the CAPP-12TB axion dark matter
experiment~\cite{CAPP} which is expected to be sensitive to the
Dine-Fischler-Srednicki-Zhitnitskii (DFSZ) axions~\cite{DFSZ} with
help from other experimental parameters given in
Sect.~\ref{SECT:VALID} and the DAQ parameters in
Sect.~\ref{SECT:DAQPARM}.
\section{Fast digitizer}\label{SECT:FADC}      
The fast digitizer for this work is the M4i.4470-x8 from Spectrum
Instrumentation GmbH~\cite{SPECTRUM}. The main features of the
analog-to-digital converters (ADCs) are a 16-bit amplitude and a
maximum sampling rate of 180 MSamples/s, where the former is
responsible for the amplitude resolution and the latter for the time
resolution, respectively. An external reference clock at 10 MHz has
been used for the ADC operation. Two additional important capacities
of the digitizer, an on-board memory of 2 GSamples and a transfer
speed of $\sim$3 GB/s through the PCI Express x8 Gen2 interface, which
can provide high efficiency on-board signal processing with proper DAQ
parameters, enable this work.
\subsection{Trigger}
Since no specific events that could be detected as trigger events
exist in the case of axion haloscope experiments, the trigger was
activated forcibly or, equivalently, the trigger condition was set to
be satisfied in all events.
\subsection{DAQ modes}
Standard (STD) and First In, First Out (FIFO) are the two major options
for the acquisition modes of the digitizer. Both modes include
processes for allocating the buffer on the digitizer memory, sampling,
and transferring the data on the digitizer memory to the PC
memory. The difference between the two modes is about when and how the
data are written and transferred to the buffers.
\subsubsection{STD mode}
The STD mode process can be summarized as follows.
\begin{enumerate}
    \item Define the total memory on the digitizer to be used.
    \item Start the digitizer (acquisition) and force the trigger.
    \item Wait until the acquisition is completed.
    \item Define a buffer on the PC side (transfer buffer) for the data to be transferred into.
    \item Transfer the data from the digitizer to the PC.
\end{enumerate}

For our digitizer in this work, starting the digitizer is practically
equivalent to starting the DAQ process. The acquisition of pretrigger data
begins when the digitizer is started, and continued until the trigger
is activated. After the trigger, the acquisition of posttrigger data
is controlled by the digitizer, independently of system usage on the
PC side. Completion of the acquisition can be recognized by writing a
relevant register to the digitizer, which is released after all the
data are acquired. This process is necessary in order to minimize the
time gap between the end of the acquisition and the start of the data
transfer. The data then can be transferred from the digitizer to the
PC and post processing of the data can be performed.
\subsubsection{FIFO mode}
The FIFO mode process is summarized next.
\begin{enumerate}
    \item Define the total memory on the digitizer to be used.
    \item Start the digitizer (acquisition) and force the trigger.
    \item Define the transfer buffer.
    \item Start the data transfer.
    \item Wait until the particular size of the data (referred to as notify size) is acquired and transferred.
    \item Loop the procedures 4 and 5 until the desired amount of data is acquired. Stop the acquisition and transfer explicitly with the user command.
\end{enumerate}

In the FIFO mode, the data transfer can be started at any moment,
even before the sampling has started, once the necessary buffers
are defined. Since the data transfer speed is faster than the sampling
rate, the data transfer can be done almost simultaneously with
sampling, hence no additional waiting for data transfer is needed.

While the data transfers, an interrupt occurs each time the notify
size of the data is transferred. After the interrupt, the transferred
data can be post processed or saved to disk. Starting the
post processing at each interrupt reduces the memory
usage on the PC side significantly, particularly in the axion
haloscope DAQ system, since at the end only the averaged spectrum in a
particular frequency range will be saved as the final output. After
that the acquisition and transfer can either be stopped, or continued
for the next set of data. For the latter case it is necessary to release the
space on the digitizer buffer occupied by the data already
transferred, so that it can be used to write the new data when the
buffer is full. This is particularly necessary when the total size of
the data to be acquired is larger than the installed memory on the
digitizer. For such procedures the buffer on the digitizer side is
initially page-aligned, and the transfer buffer defined on the PC
memory should be programmed to be page-aligned as well.

The choice of notify size is restricted to multiples of the operating
system's page size. Since the acquisition and transfer can only be
stopped after the notify size of the data is transferred, there can be
unnecessary additional sampling (oversampling) if the size of the
desired number of samples is not divisible by the notify size. In
practice, the additional time due to oversampling is negligible, since
the total amount of data is usually significantly larger than the minimum
notify size, while the maximum size of data in addition to what is
required is less than the notify size. As an example, the total number
of samples to be acquired is 225M for 10 seconds of sampling, with a
sampling rate of 22.5 MSamples/s, while a typical minimum notify size is
4096 bytes, corresponding to 2048 samples. This results in at most 1664
additional samples, corresponding to 74 $\mu$s of time.
\subsection{Parameters}\label{SECT:DAQPARM}
The typical intermediate frequency (IF) of 10.7 MHz can be chosen at
the heterodyne receiver for the CAPP-12TB experiment. With a sampling
rate of 22.5 MSamples/s, the Nyquist frequency is 11.25 MHz which can
result in a power spectrum span of 1.1 MHz with the IF as the central
frequency. The span of 1.1 MHz is much wider than the relevant
CAPP-12TB parameters, cavity bandwidths of $\sim$30 kHz and virialized
axion~\cite{turner} signal windows of $\sim$3 kHz, and thus meets the
criteria of CAPP-12TB axion haloscope searches.

With a resolution bandwidth (RBW) of 20 Hz, data from this process
can be used for nonvirialized axions~\cite{HR-AXION} whose signal
windows are at most 20 Hz as well as virialized axions mentioned
above. This narrow RBW is also more effective for getting rid of IF
and radio frequency (RF) interference as the narrow spikes in each
power spectrum~\cite{HAYSTAC-PRD} of the virialized axion dark matter
searches that is what we are primarily after.

We chose a dynamic range of $\pm$500 mV, which limits very high power
electromagnetic interference backgrounds and provides a power
resolution of about 0.5\% with a 16-bit amplitude resolution of the
digitizer and expected background power level of $\mathcal{O}(\mu$W).
Table~\ref{TAB:DIGI-PARM} lists the digitizer parameters used in this
work.
\begin{table}[h]
  \centering
  \begin{tabular}{|c|c|c|c|c|}\hline
    sampling    & spectrum &RBW & number of ADC data &dynamic\\ 
        rate    & span     &    & per power spectrum &range\\ \hline
    22.5 MSamples/s & 1.1 MHz       &20 Hz & 1125000 &$\pm$500 mV\\ \hline
  \end{tabular}
  \caption{Parameters for the digitizer.}      
  \label{TAB:DIGI-PARM}
\end{table}
\section{Axion haloscope DAQ system}
In general, the main data for axion haloscope searches consists of
continuous RF signals resulting in power spectra at the end, where the
power spectra can be acquired from online or offline FFTs. Online FFT
can be done with a commercial spectrum analyzer.
However, the rather low $\epsilon_{\rm DAQ}$ of 47\% obtained with the
spectrum analyzer R\&S$^{\tiny\circledR}$FSV4~\cite{RSH} in the
CAPP-8TB experiment~\cite{CAPP-8TB-NIM} mentioned above, resulted mainly
from the online FFT of the spectrum analyzer.
Although offline FFT can be realized with an extra conventional
storage unit to save the ADC data for a chosen sampling rate of 22.5
MSamples/s in this work, here we will employ the online FFT as done in
Ref.~\cite{HAYSTAC-NIM} using the full advantages of the digitizer, an
on-board memory of 2 GSamples and a transfer speed of $\sim$3
GB/s. Our recent development of a field-programmable gate array
realized a realtime DAQ system including the online FFT, albeit with a
fixed RBW of 100 Hz and a power spectrum span of 500
kHz~\cite{CAPP-FPGA}. The DAQ system developed in this work is more
flexible in the power spectrum parameters, RBW and span, in addition
to the image rejection using two parallel DAQ channels.

\section{Single channel DAQ process}
\subsection{Parallel data processing and sampling}
The challenging part of our DAQ system using the fast digitizer is the
post processing of the data, which includes unit conversion, online
FFT, averaging, and writing the power spectra (ROOT~\cite{ROOT} format
in this work) to disk storage. The online FFT dominates the post
processing time.
In most cases axion haloscope experiments require data at different
resonance frequencies, because the axion mass is unknown.
Also for various reasons the data at each resonance
frequency can be divided into several subsets at different timestamps.
In such cases, the post process can be performed in parallel while the
next data is sampled. Python's multiprocessing
module~\cite{PYTHON_MULTI} was used to demonstrate such a scenario.

The essential parts of the post processing including writing to disk
are defined separately, then integrated into a single function. The
function is then called in a child process spawned by the main process
at the end of the first DAQ cycle. The DAQ cycle ends as the child
process starts, and the main process controls the digitizer to start a
new DAQ cycle while the child process is running. After the new
process ends the main process lets the child process join, then spawns
more child processes for post processing the data from the new
DAQ cycle. These procedures are repeated until the last DAQ cycle, and
the main process ends before the post processing of data from the last
DAQ cycle starts. For the detailed $\epsilon_{\rm DAQ}$ studies given
in Sect.~\ref{SECT:eDAQ}, the main process ends after the post
processing of data from the last DAQ cycle ends.

In the child process, the data arrays are, at first, converted into a
two dimensional array with each row containing a single spectrum.
Then two child processes are again spawned, one for FFT and
the other one for averaging the spectra and saving into a file. During
the FFT process the data in each row are converted into voltage, and
Fourier transformed.
The output of the FFT is normalized and then converted into a single
power spectrum. The power spectrum is finally put into a queue defined
globally at the beginning of the main process. The other process
receives the power spectra from the queue whenever it is filled, adds
them together in a new buffer, divides it by the number of spectra
after the last power spectrum is added, then saves it in the data
file. Finally the two child processes merge with their parent
processes, and again merge with the main process.

The main process waits until the child process from a previous DAQ
cycle merges with its parents. This means the next DAQ cycle might be
delayed if the total time necessary for the post process exceeds the
DAQ cycle time segment explained below. The total time for post
processing can be reduced in this case by employing several parallel
jobs in multi-thread CPUs.

\begin{figure}[h]
  \centering
  \includegraphics[width=1.0\textwidth]{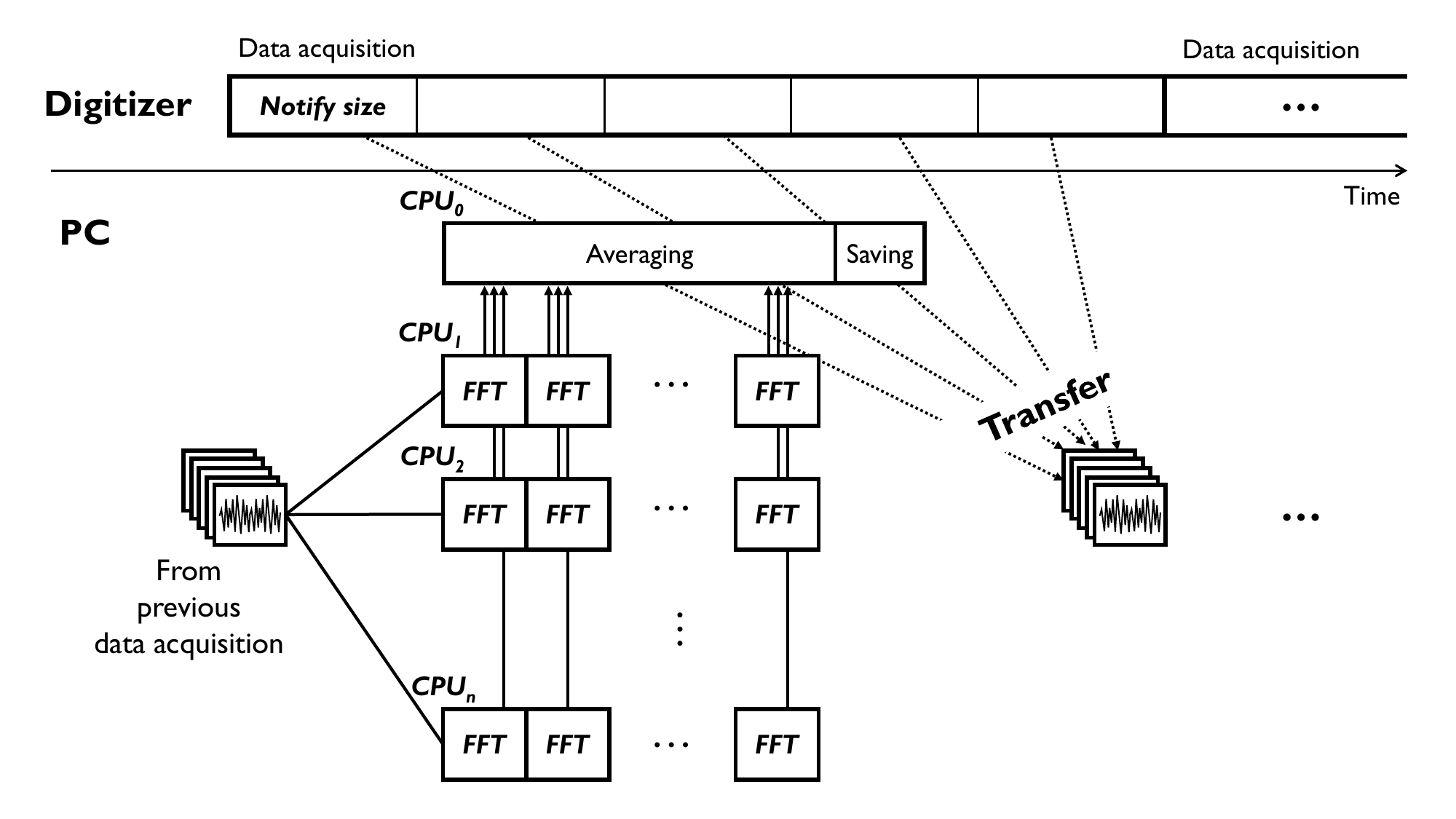}
  \caption{Schematic of the DAQ process.}
  \label{FIG:DAQFLOW}
\end{figure}

With multi-thread CPUs, each FFT process can be shortened by
multi-threading, or the overall tasks can be divided and done by
several threads in parallel. Considering the large number of
iterations and the fairly short time for each iteration, the latter is
in practice more efficient. Since the FFT processes are already being
performed in parallel and automatically assigned to another thread,
using more threads can be done simply by dividing the tasks equally
and spawning more FFT processes for assigning the divided
tasks. Figure~\ref{FIG:DAQFLOW} shows the schematic of our DAQ
process described above.
\subsection{Optimization of the DAQ cycle time segment}
Here we address how the DAQ cycle time segment was optimized with our
DAQ PC which has a 16-thread CPU whose maximum clock speed is 5.3 GHz
according to the manufacturer~\cite{INTEL}. We found the optimum cycle
time by varying the number of FFTs or, equivalently, the number of
power spectra, per thread.
\begin{figure}[h]
  \centering
  \includegraphics[width=1.0\textwidth]{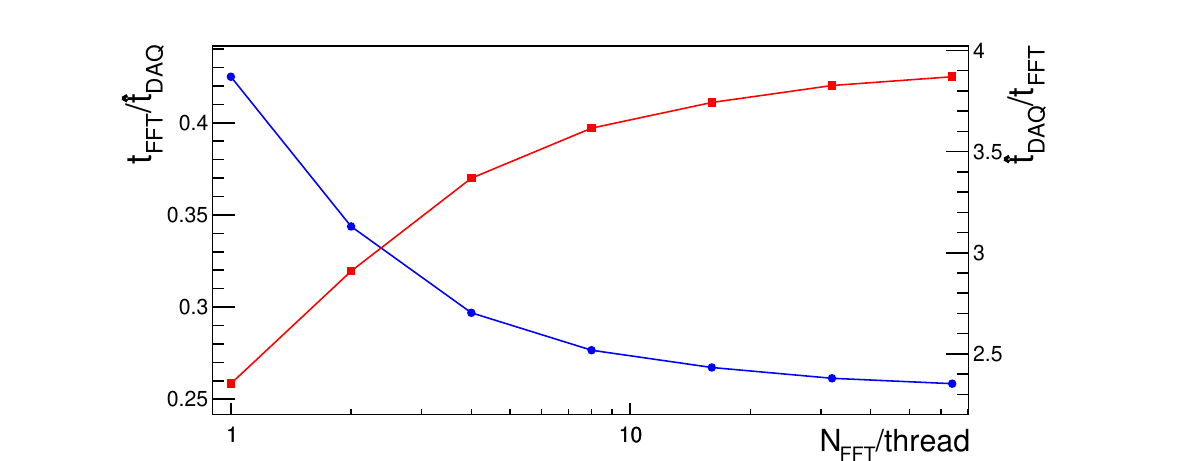}
  \caption{Time ratio of FFT to DAQ cycle (blue circles) and its reciprocal
    (red rectangles) as a function of number of FFTs per thread with
    the STD mode, where the latter can be read as the speedup
    depending on the parallel portion for a given number of
    threads~\cite{AMD}.}
  \label{FIG:FFTvsDAQ}
\end{figure}
Figure~\ref{FIG:FFTvsDAQ} shows that a time ratio of FFT to DAQ cycle
below 30\% can be obtained by choosing the number of FFTs per thread
to be greater or equal to 4. Our choice of the number of FFTs per
thread was 8. From 128 FFTs (or power spectra) with 16 threads,
therefore, our DAQ cycle time segment $\hat{t}_{\rm DAQ}$ was 6.4 s,
corresponding to a data size of 288 MB for a single channel.
\subsection{DAQ efficiency}\label{SECT:eDAQ}
From the radiometer equation~\cite{DICKE}, the experimentally achieved
SNR with an $\epsilon_{\rm SNR}$ of 100\%~\cite{Anal-JHEP} is
\begin{equation}
  {\rm SNR_{achieved}}
  =\frac{P^{a\gamma\gamma}_a}{\sigma_{P_n}}
  =\frac{P^{a\gamma\gamma}_a}{P_n}\sqrt{b_a t_{\rm DAQ}}
  =\frac{P^{a\gamma\gamma}_a}{P_n}\sqrt{N},
  \label{EQ:DICKE}
\end{equation}
where $\sigma_{P_n}$ is the $P_n$ fluctuation, $t_{\rm DAQ}$ is the
DAQ cycle time, and $N$ is the number of power spectra, which shows
that the axion dark matter search sensitivity can be improved by
increasing the DAQ cycle time or, equivalently, the number of power
spectra, by increasing the number of DAQ cycles. Therefore,
$t_{\rm DAQ}$ is equal to $n_{\rm iter}\hat{t}_{\rm DAQ}$ and our
$\epsilon_{\rm DAQ}$ depends on $n_{\rm iter}$ with our DAQ algorithm
explained above, where $n_{\rm iter}$ is the number of iterations for
each resonant frequency scan.
In our DAQ algorithm developed in this work, the first iteration time
includes the digitizer starting ($t_{\rm digitizer~starting}$) and DAQ
cycle, while the second includes the next DAQ cycle and online FFTs
with ADC data from the first iteration, where online FFTs are executed
in parallel with the current DAQ cycle. Subsequent iterations are
identical to the second iteration, until the last iteration which is
like the second one except for the additional online FFTs using
ADC data from the iteration ($t_{\rm last~FFTs}$).
Therefore, the DAQ efficiencies of the first and last iterations are
lower because of the digitizer starting and the last online FFTs,
respectively. The two inefficiency sources, however, become negligible
as the $n_{\rm iter}$ increases. Even without the $n_{\rm iter}$
increasing, the inefficiency from the last online FFTs can be in
practice ignored because the next resonant frequency scan can start
right after the final iteration of data sampling, and the
corresponding online FFTs can be finished during the cavity tuning and
measurements before starting the new DAQ process.
\begin{figure}[h]
  \centering
  \includegraphics[width=0.9\textwidth]{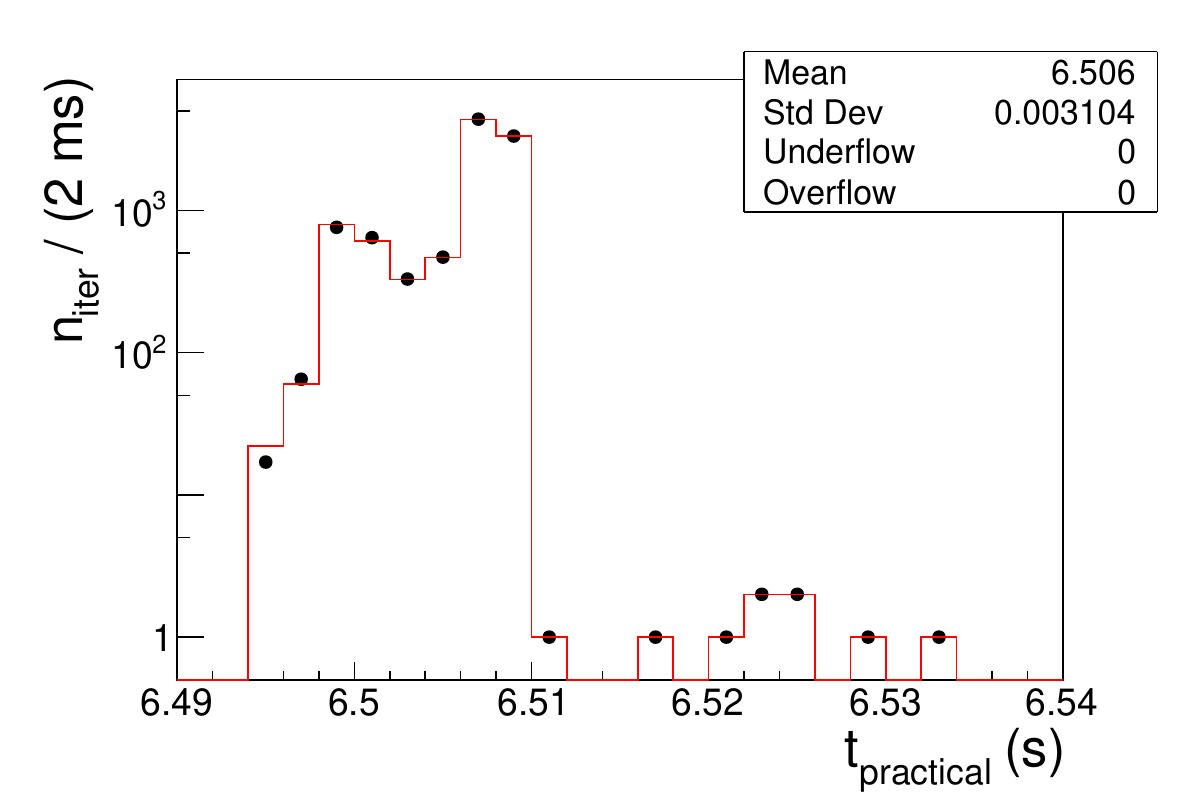}
  \caption{Timestamps from the digitizer (black circular markers) and
    DAQ PC (red histogram) with the STD mode.
    $t_{\rm practical}$ is equal to
    $t_{\rm total}-t_{\rm last~FFTs}-t_{\rm digitizer~starting}$ and
    used to estimate the $\epsilon_{\rm DAQ}$ with a practical number of
    iterations for a targeted experimental sensitivity.    
    $t_{\rm total}$ is the total time including
    $t_{\rm digitizer~starting}$ and $t_{\rm last~FFTs}$ from the
    first and last iterations, respectively, as well as $t_{\rm DAQ}$,
    where $t_{\rm digitizer~starting}$ and
    $t_{\rm last~FFTs}$ are explained the text above.}  
  \label{FIG:TIMESTAMP}
\end{figure}

The timestamp of the digitizer has limited usage, but turns out to be
consistent with that of DAQ PC as shown in
Fig.~\ref{FIG:TIMESTAMP}. This enables us to use handy timestamps from
the PC.
\begin{table}[h]
  \centering
  \begin{tabular}{|r|l|}\hline
$\epsilon_{\rm DAQ}$ with $n_{\rm iter}=10000$                 &conditions  \\ \hline
$\frac{t_{\rm DAQ}}{t_{\rm total}}=98.37\%$                    &with the last FFTs ($t_{\rm last~FFTs}$) and digitizer starting   \\ 
$\frac{t_{\rm DAQ}}{t_{\rm total}-t_{\rm last~FFTs}}=98.38\%$  &with the digitizer starting ($t_{\rm digitizer~starting}$)        \\ 
$\frac{t_{\rm DAQ}}{t_{\rm total}-t_{\rm last~FFTs}-t_{\rm digitizer~starting}}=98.39\%$  &practical or, equivalently, without $t_{\rm last~FFTs}$ and $t_{\rm digitizer~starting}$    \\ \hline
  \end{tabular}
  \caption{$\epsilon_{\rm DAQ}$ of the STD mode with $n_{\rm iter}$ of
    10000 depending on the conditions given in the second column.}  
  \label{TAB:eDAQ}
\end{table}
\begin{figure}[h]
  \centering
  \includegraphics[width=0.9\textwidth]{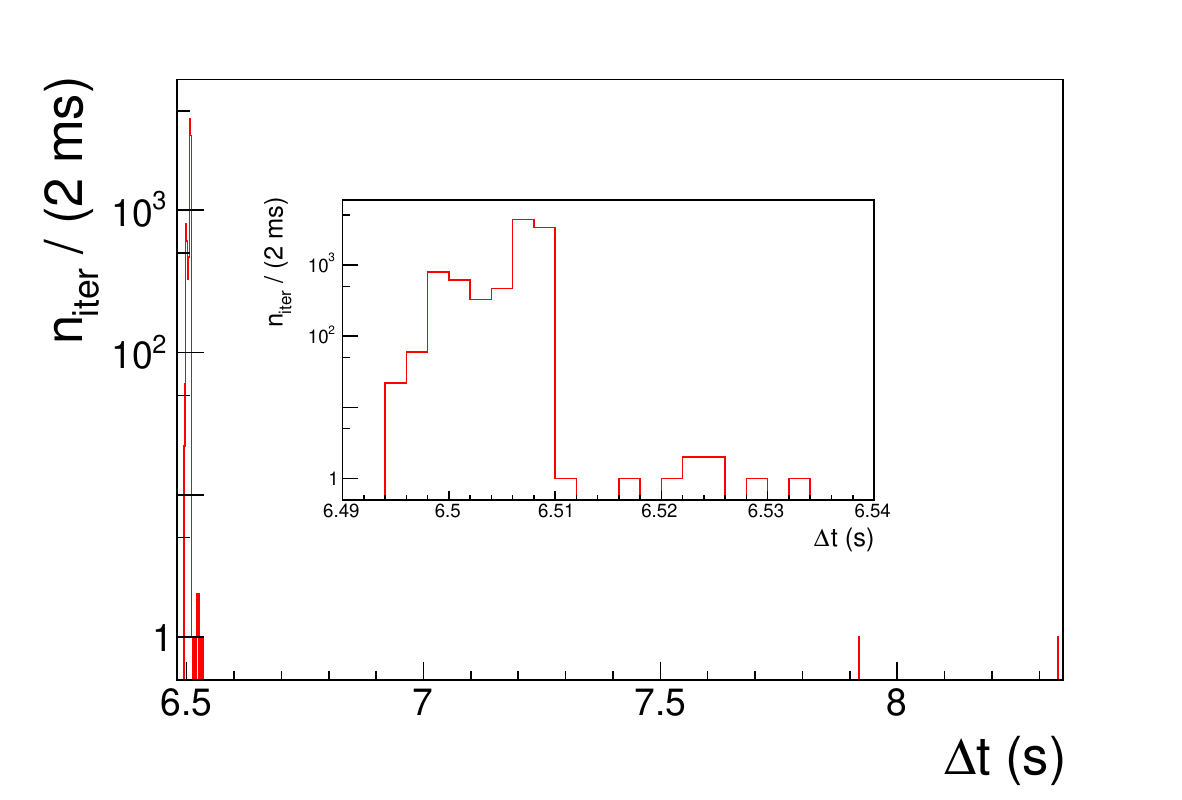}
  \caption{Distribution of $\Delta t$ with the STD mode, where the
    inset shows that $\Delta t$ around 6.5 s and is equivalent to the
    histogram in Fig.~\ref{FIG:TIMESTAMP}. Two entries $\Delta t$
    around 7.9 and 8.3 s correspond to the first and last iterations,
    respectively.}  
  \label{FIG:eDAQ_offtime}
\end{figure}
With an $n_{\rm iter}$ of 10000 and the STD mode, Table~\ref{TAB:eDAQ}
shows the $\epsilon_{\rm DAQ}$ depending on the conditions,
and Fig.~\ref{FIG:eDAQ_offtime} shows the distribution of the
timestamp differences between two consecutive iterations
$\Delta t$. As mentioned above, $\epsilon_{\rm DAQ}$ with a large
enough $n_{\rm iter}$ is independent of the two significantly
inefficient iterations shown in Fig.~\ref{FIG:eDAQ_offtime}.
\begin{figure}[h]
  \centering
  \includegraphics[width=0.9\textwidth]{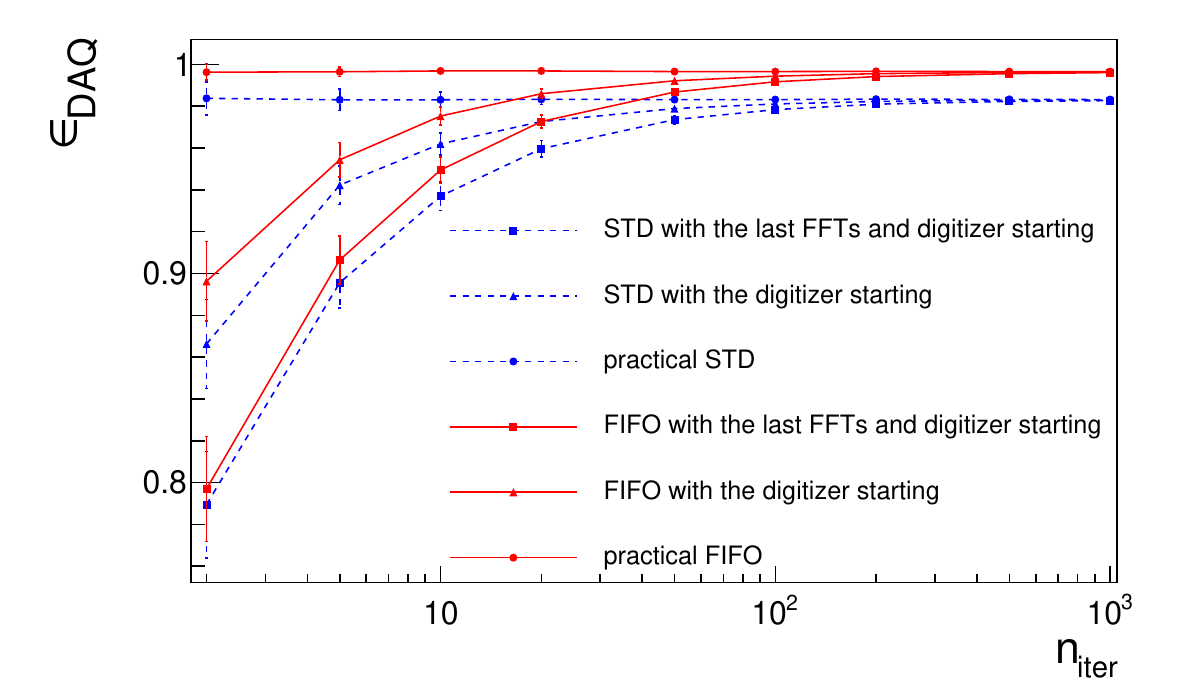}
  \caption{DAQ efficiencies with the binomial errors as a function of
    the number of iterations. The blue dashed lines (red solid lines)
    with rectangles, triangles, and circles show the efficiencies
    employing the STD (FIFO) mode depending on the conditions listed
    in Table~\ref{TAB:eDAQ}, respectively.}  
  \label{FIG:eDAQ2}
\end{figure}
Figure~\ref{FIG:eDAQ2} shows the DAQ efficiency as a function of
the number of iterations for both the STD and FIFO modes of the
digitizer. The efficiencies are already higher than 98\% and 99\%
after $\sim$200 iterations ($N$ of $\sim$26000) or $\sim$22 minutes
for the STD and FIFO modes, respectively, and then converge to 98.4\%
and 99.7\% with further iterations.
\subsection{Validation of the power fluctuations}\label{SECT:VALID}
\begin{figure}[h]
  \centering
  \includegraphics[width=0.9\textwidth]{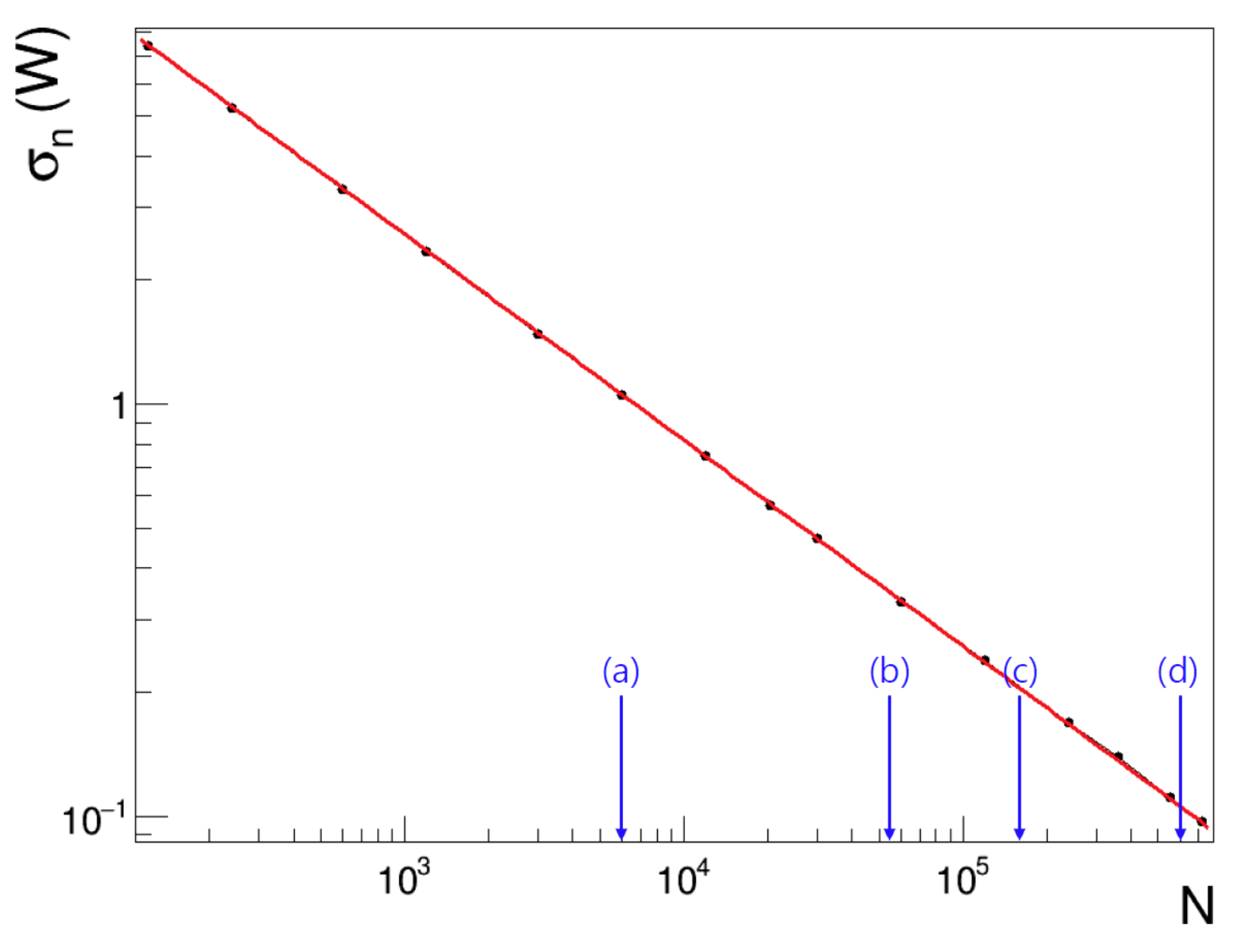}
  \caption{Noise power fluctuation $\sigma_n$ with an arbitrary
    normalization as a function of the number of power spectra $N$,
    where the dots (black) are the measurements and line (red) is the
    expected from Eq.~(\ref{EQ:DICKE}). The arrows (a), (b), (c), and
    (d) indicate the numbers of power spectra to meet
    an SNR$_{\rm achieved}$ of 5 for $T_n$ of 0.1, 0.3, 0.5 and 1 K,
    respectively, for the $g_{a\gamma\gamma}$ in
    Table~\ref{TAB:CAPP-12TB-PARM}.}  
  \label{FIG:DICKE}
\end{figure}

We checked whether the noise power fluctuations from our data with the
DAQ process follow Eq.~(\ref{EQ:DICKE}) using a 50 $\Omega$
termination input, where the fluctuations are the axion haloscope
resolutions when they are measured through a full signal chain, i.e.,
a cavity, preamps, and digitizer or spectrum analyzer at the
end. Figure~\ref{FIG:DICKE} shows the noise power fluctuations
depending on the number of power spectra, and they show good agreement
with Eq.~(\ref{EQ:DICKE}) as it should be, which is a fundamental
requirement of the DAQ system for axion dark matter searches.
Table~\ref{TAB:CAPP-12TB-PARM} shows the expected CAPP-12TB
experimental parameters for the resonant frequency $\nu=$ 1 GHz, where
the coupling strength $g_{a\gamma\gamma}$ corresponds to the DFSZ
axion and $\beta$ denotes the cavity mode coupling to the load.
\begin{table}[h]
  \centering
  \begin{tabular}{|c|c|c|c|c|c|c|c|}\hline
  $g_{a\gamma\gamma}$ &$B$ &$V$ &$C$ &$Q_L$ &$\beta$ &$\nu$ &$b_a$ \\ \hline
  5.64$\times10^{-16}$ GeV$^{-1}$ &10.85 T &30 L &0.56 &35000 &2 &1 GHz &3200 Hz \\ \hline
  \end{tabular}
  \caption{Expected CAPP-12TB experimental parameters for the resonant
    frequency $\nu=$ 1 GHz. The coupling $g_{a\gamma\gamma}$ is given
    in natural units.}  
  \label{TAB:CAPP-12TB-PARM}
\end{table}
Given the experimental parameters listed in
Table~\ref{TAB:CAPP-12TB-PARM}, Fig.~\ref{FIG:DICKE} also shows the
necessary number of power spectra to be sensitive to an
SNR$_{\rm achieved}$ of 5 for the DFSZ axions, depending on the noise
temperature. Accordingly, the DAQ efficiencies ignoring the last FFTs
are higher than 99.2\% with a noise temperature higher than 0.1 K in
the FIFO mode. This noise power fluctuation study validates that the
data from this DAQ process can be sensitive to the DFSZ axion even
with a noise temperature of 1 K, if it goes with the CAPP-12TB
experiment. Note that the statistics for this validation study was set
to match the CAPP-12TB experiment, but the validation of our DAQ
process can be extended easily to other experiments depending on the
experimental parameters.
\section{Image rejection DAQ system}
Equation~(\ref{EQ:IQX_RF}) shows the RF signals that contribute to the
IF signals in typical axion haloscope searches employing a heterodyne
receiver, where the first term on the right-hand side is the signal
$s(t)$ at $f_{\rm RF}$ and the second is the image background $b(t)$ at
$f_{\rm IM}$.
\begin{equation}
  v_{\rm RF}(t)=s(t)\cos(\omega_{\rm RF}t+\phi)+b(t)\cos(\omega_{\rm IM}t+\phi),
 \label{EQ:IQX_RF}
\end{equation}
where $\omega=2\pi f$ and $\phi$ is a phase.
Adopting $\omega_{\rm IF}=\omega_{\rm RF}-\omega_{\rm LO}$, then
$\omega_{\rm IM}=\omega_{\rm RF}-2\omega_{\rm IF}$, where LO stands for a local
oscillator.

Downconverting $v_{\rm RF}(t)$ using an RF mixer and signal generator as
an LO results in the IF signals expressed by Eq.~(\ref{EQ:IQX_IF}).
\begin{equation}
  v_{\rm IF}(t)=\frac{a_{\rm LO}}{2}(s(t)\cos(\omega_{\rm IF}t+\phi)+b(t)\cos(\omega_{\rm IF}t-\phi)),
 \label{EQ:IQX_IF}
\end{equation}
where $a_{\rm LO}$ is the LO amplitude and the second term on the
right-hand side is the unwanted image background at $f_{\rm IF}$, which
can be rejected by hardware~\cite{CAPP-8TB-NIM, ADMX-NIM} or
software~\cite{HAYSTAC-NIM}.
In this work, we employed a software-based image rejection using a
PolyPhase IQ mixer QD0622B~\cite{POLYPHASE}, and both DAQ channels on
the digitizer. 
$V^{\rm I}_{\rm IF}(f)$ and $V^{\rm Q}_{\rm IF}(f)$ are the complex FFT outputs of the
ADC data from the I and Q channels of the mixer,
respectively. One, then, can reject the image background by
constructing Eq.~(\ref{EQ:IQX_X}) with real and imaginary parts of
$V^{\rm I}_{\rm IF}(f)$ and $V^{\rm Q}_{\rm IF}(f)$.
\begin{equation}
  V_{\rm IF}(f)=(\Re(V^{\rm I}_{\rm IF}(f))+\Im(V^{\rm Q}_{\rm IF}(f)))+i(\Im(V^{\rm I}_{\rm IF}(f))-\Re(V^{\rm Q}_{\rm IF}(f))).
 \label{EQ:IQX_X}
\end{equation}
\begin{figure}[h]
  \centering
  \includegraphics[width=0.9\textwidth]{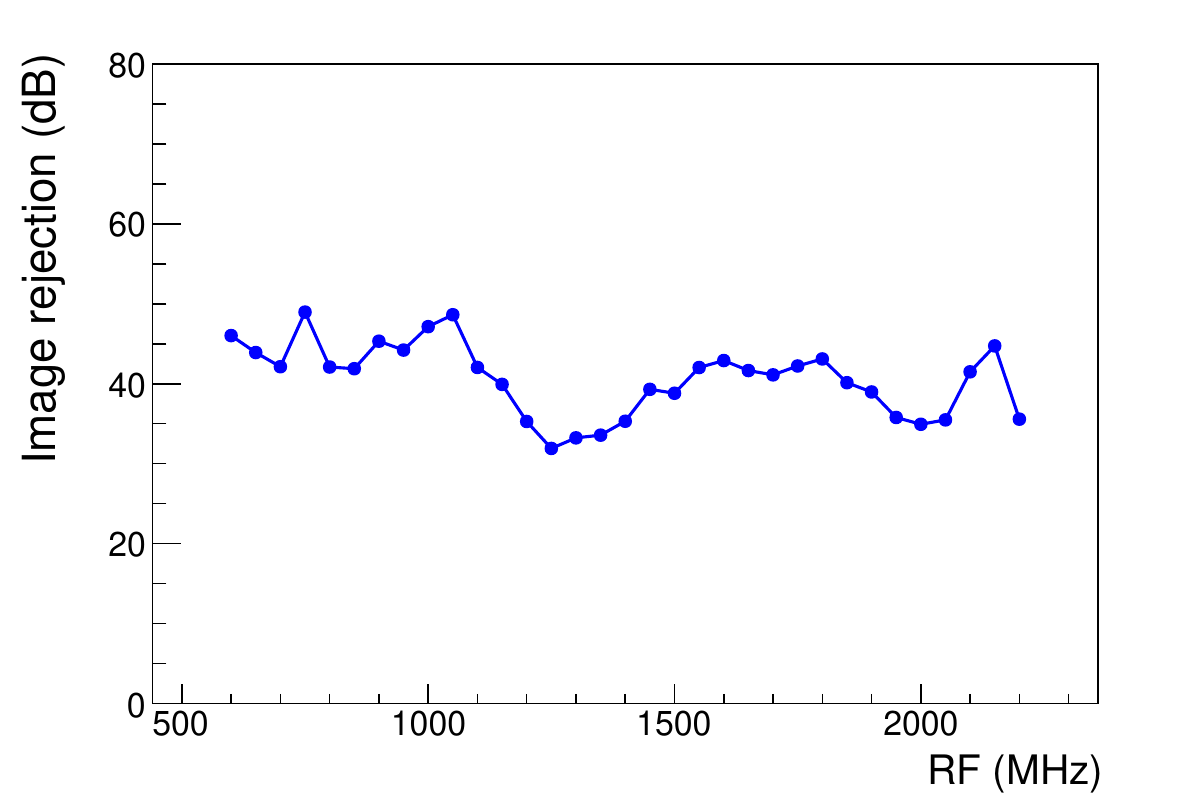}
  \caption{Image rejection in dB as a function of the RF frequency.}  
  \label{FIG:IR_IQX}
\end{figure}
\begin{figure}[h]
  \centering
  \includegraphics[width=0.9\textwidth]{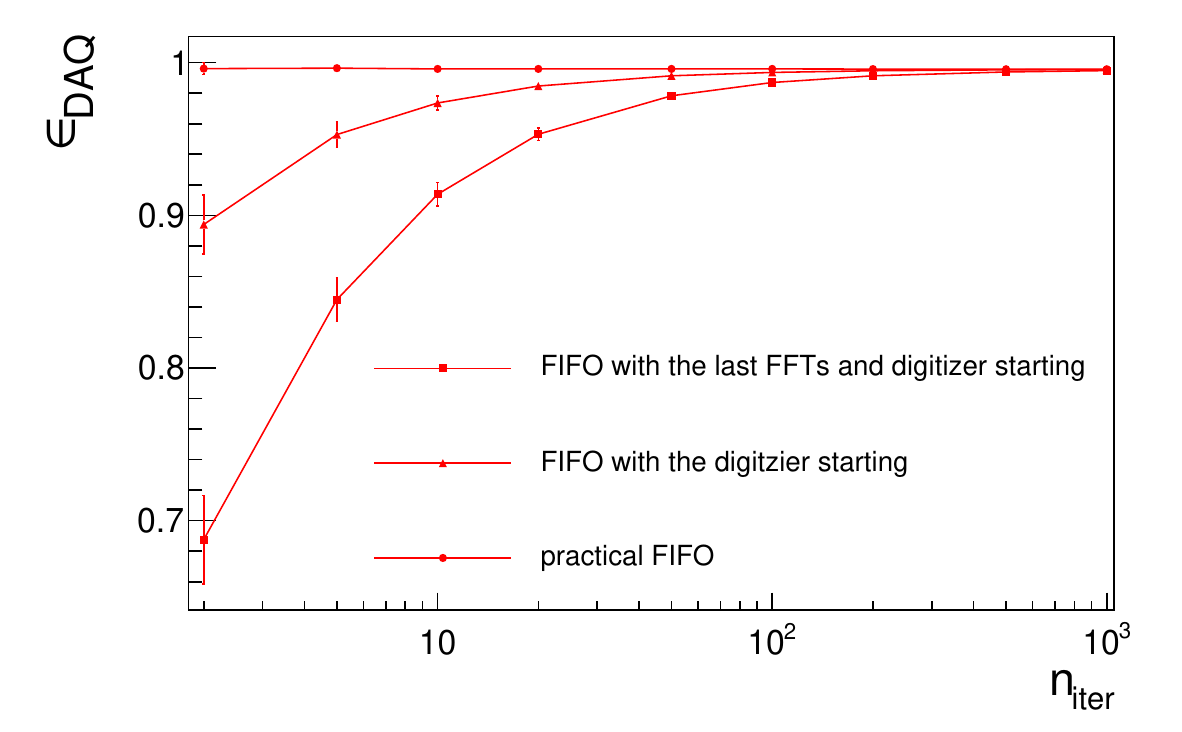}
  \caption{The same as Fig.~\ref{FIG:eDAQ2} except for the FIFO mode
    only, with two parallel DAQ channels for the image rejection.}  
  \label{FIG:eDAQ_IQX}
\end{figure}
The image rejection using Eq.~(\ref{EQ:IQX_X}) was measured over a
frequency range from 600 to 2200 MHz and Fig.~\ref{FIG:IR_IQX} shows
the results $\frac{P_{\rm input}}{P_{\rm output}}$ in dB as a function
of the RF frequency, where $P_{\rm input}$ was set to 0 dBm. Note that
$P_{\rm input}$ were injected to $f_{\rm IM}=f_{\rm RF}-2f_{\rm IF}$ and
$P_{\rm output}$ were measured at $f_{\rm IF}=10.7$ MHz. Without losing
the $\epsilon_{\rm DAQ}$ practically, as shown in
Fig.~\ref{FIG:eDAQ_IQX}, we realized a fast DAQ system equipped with
an image rejection of about 35 dB over a frequency range from 600 to
2200 MHz. Note, however, that the image rejection here can go to any
frequency with the relevant IQ mixers.

\section{Summary}
A DAQ system using a two-channel fast digitizer that can sample 16-bit
amplitudes at rates up to 180 MSamples/s was developed to improve the
figure of merit in axion haloscope searches, the scanning rate.
The achieved DAQ efficiency, including the online FFT and writing the
outputs to disk, is practically 100\%, which is a significant
improvement compared with our previous DAQ system using a commercial spectrum
analyzer~\cite{CAPP-8TB-NIM}. The limits of $g_{a\gamma\gamma}$ from
the CAPP-8TB experiment~\cite{CAPP-8TB-PRL} can be improved from about
1.00$\times10^{-14}$ to 0.83$\times 10^{-14}$ GeV$^{-1}$ by running
the same experimental period employing the DAQ system in this work and
can reach 0.79$\times 10^{-14}$ GeV$^{-1}$ by putting our previous
improvement~\cite{Anal-JHEP} together.

Using the two channels on the digitizer and an IQ mixer, our DAQ
system also has an image rejection of about 35 dB over a frequency
range from 600 and 2200 MHz, without losing the DAQ efficiency
practically, e.g., for the CAPP-12TB experiment~\cite{CAPP}. This
means the DAQ system developed in this work shows various improvements
that allow the CAPP-12TB axion dark matter experiment to be sensitive
the DFSZ axions in relevant frequency ranges.

\acknowledgments
This work is supported by the Institute for Basic Science (IBS) under
Project Code No. IBS-R017-D1-2022-a00.


\begin{thebibliography}{99}

  \bibitem{strongCP}
  G. 't Hooft, Phys. Rev. Lett, {\bf 37} (1976) 8; 
  Phys. Rev. D {\bf 14} (1976) 3432; {\bf 18} (1978) 2199(E); 
  J. H. Smith, E. M. Purcell, and N. F. Ramsey, Phys. Rev. \textbf{108} (1957) 120;
  W. B. Dress, P. D. Miller, J. M. Pendlebury, P. Perrin, and N. F. Ramsey, Phys. Rev. D {\bf 15} (1977) 9; 
  I. S. Altarev {\it et al.}, Nucl. Phys. \textbf{A341} (1980) 269. 

\bibitem{PQ}
  R. D. Peccei and H. R. Quinn, Phys. Rev. Lett. \textbf{38} (1977) 1440.
  
\bibitem{AXION}
  S. Weinberg, Phys. Rev. Lett. \textbf{40} (1978) 223;
  F. Wilczek, Phys. Rev. Lett. \textbf{40} (1978) 279.

\bibitem{CDM_LOW}
  J. Preskill, M. B. Wise, and F. Wilczek, Phys. Lett. B \textbf{120} (1983) 127;
  L. F. Abbott and P. Sikivie, Phys. Lett. B \textbf{120} (1983) 133;
  M. Dine and W. Fischler, Phys. Lett. B \textbf{120} (1983) 137.

\bibitem{OTHER_AXION_PROD}
  Fuminobu Takahashi, Wen Yin, and Alan H. Guth, Phys. Rev. D {\bf 98} (2018) 015042;  
  Peter W. Graham and Adam Scherlis, Phys. Rev. D {\bf 98} (2018) 035017.

\bibitem{SN1987}
  John Ellis and K. A. Olive, Phys. Lett. B \textbf{193} (1987) 525;
  Georg Raffelt and David Seckel, Phys. Rev. Lett. \textbf{60} (1988) 1793;  
  Michael S. Turner, Phys. Rev. Lett. \textbf{60} (1988) 1797;
  Hans-Thomas Janka, Wolfgang Keil, Georg Raffelt, and David Seckel, Phys. Rev. Lett. \textbf{76} (1996) 2621;
  Wolfgang Keil, Hans-Thomas Janka, David N. Schramm, G$\ddot{\rm u}$nter Sigl, Michael S. Turner, and John Ellis, Phys. Rev. D \textbf{56} (1997) 2419.  

\bibitem{PLANCK}
  P. A. R. Ade {\it et al.} (Planck Collaboration), Astron. Astrophys. \textbf{594} (2016) A13. 

\bibitem{sikivie}
  P. Sikivie, Phys. Rev. Lett. \textbf{51} (1983) 1415; Phys. Rev. D \textbf{32} (1985) 2988.

\bibitem{scanrate}
  L. Krauss, J. Moody, F. Wilczek, and D. E. Morris, Phys. Rev. Lett. \textbf{55} (1985) 1797.

\bibitem{EMFF_BRKO}
  B. R. Ko {\it et al.}, Phys. Rev. D \textbf{94} (2016) 111702(R).  

\bibitem{Anal-JHEP}
  S. Ahn, S. Lee,  J. Choi, B. R. Ko, and Y. K. Semertzidis, J. High Energ. Phys. \textbf{2021} (2021) 297.
  
\bibitem{RSH}
  \url{https://www.rhode-schwarz.com}.

\bibitem{CAPP-8TB-NIM}
  J. Choi, S. Ahn, B. R. Ko, S. Lee, and Y. K. Semertzidis, Nucl. Instrum. Methods Phys. Res., Sect. A \textbf{1013} (2021) 165667.

\bibitem{ADMX-NIM}
  R. Khatiwada {\it et al.} (ADMX Collaboration), \url{arXiv:2010.00169}. 

\bibitem{HAYSTAC-NIM}
  S. Al Kenany {\it et al.}, Nucl. Instrum. Methods Phys. Res., Sect. A \textbf{854} (2017) 11.

\bibitem{CAPP}
  Y. K. Semertzidis {\it et al.}, \url{arXiv:1910.11591}.

\bibitem{DFSZ}
  A. R. Zhitnitskii, Yad. Fiz. {\bf 31} (1980) 497
  [Sov. J. Nucl. Phys. \textbf{31} (1980) 260];
  M. Dine, W. Fischler, and M. Srednicki, Phys. Lett. B \textbf{140} (1981) 199.

\bibitem{SPECTRUM}
  \url{https://spectrum-instrumentation.com}.

\bibitem{turner}
  M. S. Turner, Phys. Rev. D \textbf{42} (1990) 3572.

\bibitem{HR-AXION}
  P. Sikivie, I. I. Tkachev, and Y. Wang, Phys. Rev. D \textbf{56} (1997) 1863;
  P. Sikivie, Phys. Lett. \textbf{B} 567 (2003) 1;
  K. Freese, P. Gondolo, H. J. Newberg, and M. Lewis, Phys. Rev. Lett. \textbf{92} (2004) 111301;
  K. Freese, P. Gondolo, and H. J. Newberg, Phys. Rev. D \textbf{71} (2005) 043516;
  L. D. Duffy and P. Sikivie, Phys. Rev. D \textbf{78} (2008) 063508.
  
\bibitem{HAYSTAC-PRD}
  B. M. Brubaker, L. Zhong, S. K. Lamoreaux, K. W. Lehnert, and K. A. van Bibber, Phys. Rev D \textbf{96} (2017) 123008.

\bibitem{CAPP-FPGA}
  M. J. Lee, B. R. Ko, and S. Ahn, JINST \textbf{16} (2021) T11008.

\bibitem{ROOT}
  R. Brun, F. Rademakers, Nucl. Instrum. Methods Phys. Res., Sect. A \textbf{389} (1997) 81.

\bibitem{PYTHON_MULTI}
  M.M. McKerns, L. Strand, T. Sullivan, A. Fang, M.A.G. Aivazis,
  \url{http://arxiv.org/pdf/1202.1056};
  Michael McKerns and Michael Aivazis,
  \url{https://uqfoundation.github.io/project/pathos}.

\bibitem{INTEL}
  \url{https://ark.intel.com/content/www/us/en/ark/products/212325/intel-core-i911900k-processor-16m-cache-up-to-5-30-ghz.html}.
  
\bibitem{AMD}
  Gene M. Amdahl, AFIPS Joint Spring Conference Proceedings 30
  (Atlantic City, NJ, 1967), AFIPS Press, Reston VA, pp 483-485.

\bibitem{DICKE}
  R. H. Dicke, Rev. Sci. Instrum. \textbf{17} (1946) 268.

\bibitem{POLYPHASE}
  \url{https://polyphasemicrowave.com}.

\bibitem{CAPP-8TB-PRL}
  S. Lee, S. Ahn, J. Choi, B. R. Ko, and Y. K. Semertzidis, Phys. Rev. Lett. \textbf{124} (2020) 101802.
  
\end{thebibliography}
\end{document}